\documentclass[twocolumn,trackchanges]{aastex631}

\newcommand{\nmad}{$NMAD_{loss}$}
\newcommand\mperiod[1][\rlap]{#1{\;\;\;.}}
\newcommand\mcomma[1][\rlap]{#1{\;\;\;,}}
\usepackage{graphicx}
\usepackage{caption,subcaption}
\usepackage{enumitem}

\submitjournal{ApJ}

\shorttitle{SpecPT model for Galaxy Spectra I}
\shortauthors{Pattnaik et al.}

\graphicspath{{./}{}}

\begin{document}

\title{SpecPT (Spectroscopy Pre-trained Transformer) Model for Extragalactic Spectroscopy: I. Architecture and Automated Redshift Measurement}

\author[0000-0003-3835-9898]{Rohan Pattnaik}
\affil{Laboratory for Multiwavelength Astrophysics, School of Physics and Astronomy, Rochester Institute of Technology, 84 Lomb Memorial Drive, Rochester, NY 14623, USA}

\author[0000-0001-9187-3605]{Jeyhan S. Kartaltepe}
\affil{Laboratory for Multiwavelength Astrophysics, School of Physics and Astronomy, Rochester Institute of Technology, 84 Lomb Memorial Drive, Rochester, NY 14623, USA}

\author[0009-0009-4635-9442]{Clive Binu}
\affil{Laboratory for Multiwavelength Astrophysics, School of Physics and Astronomy, Rochester Institute of Technology, 84 Lomb Memorial Drive, Rochester, NY 14623, USA}

\begin{abstract}

We introduce the Spectroscopy Pre-trained Transformer (SpecPT), a transformer-based model designed to analyze spectroscopic data, with applications in spectrum reconstruction and redshift measurement. Using the Early Data Release (EDR) of the DESI survey, we evaluate SpecPT’s performance on two distinct datasets: the Bright Galaxy Survey (BGS) and Emission Line Galaxy (ELG) samples. SpecPT successfully reconstructs spectra, accurately capturing emission lines, absorption features, and continuum shapes while effectively reducing noise. For redshift prediction, SpecPT achieves competitive accuracy, with Normalized Median Absolute Deviation (NMAD) values of 0.0006 and 0.0008, and catastrophic outlier fractions of 0.20\% and 0.80\% for BGS and ELG, respectively. Notably, SpecPT performs consistently well across the full redshift range ($0 < z < 1.6$), demonstrating its versatility and robustness. By leveraging its learned latent representations, SpecPT lays the groundwork for a foundational spectroscopic model, with potential applications in outlier detection, interstellar medium (ISM) property estimation, and transfer learning to other datasets. This work represents a first step in building a generalized framework for spectroscopic analysis, capable of scaling to the full DESI dataset and beyond.

\end{abstract}


\section{Introduction} \label{sec:intro}

Extragalactic spectroscopy is a foundational tool in astrophysics, offering insights into the evolution of galaxies and the broader cosmos. Through the analysis of galaxy spectra, we can infer a range of critical properties, including the star formation rate (SFR), metallicity (Z), ionization parameter (U), gas pressure, extinction, and other characteristics of the interstellar medium (ISM, \citealt{kewley2019understanding}). However, the extraction of vital information from these spectra for analysis is a complex and often labor-intensive process. Current methods of fitting models are not only time-consuming but also become significant bottlenecks when dealing with large datasets. Consequently, there is a pressing need for alternative methods capable of processing millions of spectra more efficiently, allowing for the rapid extraction of key information and thus accelerating the pace of astrophysical research.

Central to these studies is the accurate measurement of spectroscopic redshift, which plays a crucial role in advancing modern extragalactic astrophysics, facilitating a deeper understanding of the Universe's large-scale structure and addressing fundamental questions about its properties. Major surveys, such as 2dF \citep{colless20012df}, VIMOS's VVDS \citep{garilli2010ez}, SDSS's BOSS \citep{bolton2012spectral}, and GAMA \citep{baldry2014galaxy}, have been instrumental in providing high-quality data that significantly contribute to our progress in this field. These surveys offer valuable redshift measurements, enabling the study of various phenomena, such as the accretion and growth of supermassive black holes, the formation of baryonic matter and dark matter halos in the large-scale structure, and the properties of star-forming galaxies at different redshifts, including their dependence on environmental factors \citep{rix2004gems,martin2005galaxy}.

Estimating galaxy properties accurately from spectra relies on the essential step of transforming observed wavelengths to rest-frame wavelengths using the galaxy's redshift. Reliable redshift measurements are important for calculating accurate distances to objects. Although redshifts can be measured through photometric or spectroscopic observations, the latter offer a higher level of precision and accuracy. Photometric redshifts (photozs) are obtained through the fitting of models to the Spectral Energy Distributions (SED) of galaxies, but they tend to be over an order of magnitude less precise compared to spectroscopic measurements \citep{ilbert2008cosmos}.

Spectroscopic measurements, despite being more resource-intensive, offer redshift estimates with precision on the order of $10^{-3}$ for a spectral resolution of R $\sim$ 200 \citep{le2005vimos}. Such high precision is particularly valuable for generating environmental maps of galaxies in the local and early Universe and quantifying the effects of environment on galaxy properties across different redshifts. However, obtaining the redshift of a galaxy from its observed spectrum is not a trivial task. The process involves visually inspecting each individual spectrum to identify and fit spectral features, such as emission and absorption lines, or employing cross-correlation techniques with galaxy or stellar templates at varying redshifts. Traditionally, astronomers have used tools like IRAF's \texttt{rvsao} \citep{kurtz1998rvsao} for cross-correlation using $\chi^{2}$ fitting. Both methods — manual visual inspection and automated cross-correlation techniques — come with significant drawbacks: the former is time-intensive and reliant on human effort for consistency, while the latter can struggle with complex or noisy spectra, leading to inaccuracies in redshift determination.

To streamline this process, attempts have been made to automate redshift estimation for spectroscopic data from large-scale surveys, including VIMOS's VVDS \citep{garilli2010ez}, SDSS's BOSS \citep{bolton2012spectral}, and GAMA \citep{baldry2014galaxy}, to name a few. However, existing automated methods still rely on the cross-correlation technique, which can be slow and prone to errors. Advancements in automating redshift measurement techniques are necessary to fully leverage the potential of spectroscopic data and further enhance our understanding of the Universe's large-scale structure and galaxy properties.

A recent and promising development in redshift measurement techniques involves the application of deep learning and machine learning algorithms. These methods have shown extensive success in measuring photozs (e.g., \cite{collister2004annz,wadadekar2004estimating,gerdes2010arborz,way2012can,carrasco2013tpz,hogan2015gaz,hoyle2016measuring,schuldt2020photometric}). Machine learning has proven to be a valuable tool in easing the computational burden and providing a deeper understanding of the parameter space that influences galaxy photozs. These algorithms leverage vast training datasets, learning intricate relationships between photometric features and true redshifts, thereby enabling efficient and accurate redshift estimates for large samples of galaxies.

Despite the success of ML algorithms in accurately estimating photozs, little work had been done until recently on directly applying ML to measure redshifts from spectra. Published studies like \cite{stivaktakis2019convolutional} and \cite{zhou2021spectroscopic} were among the first to demonstrate that CNNs can successfully retrieve accurate spectroscopic redshift measurements and provide confidence scores. To provide a confidence score, both methods transformed redshift estimation into a classification task by dividing the range into discrete bins. However, increasing the precision beyond photozs requires a high number of bins, leading to increased network complexity. This approach worked well for limited-size input spectra, but may struggle with larger, noisier spectra. Additionally, dividing spectra into bins constrains the maximum achievable redshift precision. However, these studies offered promising proof of concept, paving the way towards the use of deep learning methods for spectroscopic analysis.

Recent advancements in redshift measurement have seen the adoption of more sophisticated deep learning methods, particularly those leveraging state-of-the-art Autoencoder and Transformer architectures. For instance, AstroCLIP \citep{lanusse2023astroclip, parker2024astroclip} introduces a cross-modal foundation model that embeds both galaxy images and spectra into a shared latent space, enabling versatile downstream tasks such as photometric redshift estimation and galaxy property prediction. This approach exemplifies the potential of transformer-based architectures in handling diverse types of astronomical data within a unified framework.

Similarly, GaSNet-II \citep{zhong2023galaxy} applies a deep learning framework to spectroscopic data, achieving high classification accuracy and precise redshift predictions across various datasets, including those from Sloan Digital Sky Survey (SDSS) and the Dark Energy Spectroscopic Instrument (DESI). The efficiency and accuracy demonstrated by GaSNet-II highlight the growing capability of deep learning models in processing large-scale spectroscopic surveys in real-time, which is critical for the ongoing and future demands of astronomical research.

Additionally, the SPENDER architecture of \cite{melchior2023autoencoding} exemplifies the application of Autoencoders in galaxy spectra analysis, offering robust spectral reconstructions and innovative outlier detection mechanisms \citep{liang2023autoencoding,liang2023outlier}. These developments underscore the rapid progress being made in utilizing advanced deep learning techniques for spectroscopic data, pushing the boundaries of what can be achieved in the field.

In this paper, we take a significant step toward developing a foundational model for spectroscopic analysis and measuring redshifts directly from input spectra. Building upon recent advancements in deep learning, particularly the use of Transformer and Autoencoder architectures, we introduce SpecPT, a Transformer-based model designed to handle the complexities of diverse spectroscopic datasets. Our approach is motivated by the need for a flexible and scalable solution that can be applied across various spectroscopic datasets, a gap that current methods, despite their successes, have yet to fully address.

To achieve this, we leverage the extensive DESI datasets, particularly the Bright Galaxy Survey (BGS) and the Emission Line Galaxy (ELG) samples, to train and test our model. By using these datasets, which have already demonstrated their value in the development of cutting-edge algorithms like those in AstroCLIP and GaSNet-II, we aim to create a framework that not only provides accurate redshift estimates but is also adaptable to any spectroscopic dataset.

The paper is structured as follows: Section \ref{sec:data} introduces the DESI data used for training and testing the SpecPT model. Section \ref{sec:SpecPT} describes the architecture and operation of the SpecPT model, focusing on its transformer-based design and the optimization methods employed to enhance its performance across various spectroscopic datasets. Section \ref{sec:results} presents the results of training and testing SpecPT on the DESI catalogs, offering a thorough analysis of its effectiveness. In Section \ref{sec:discussion}, we discuss these results, followed by a summary of key findings, conclusions, and directions for future work in Section \ref{sec:summary}.

\section{Data}
\label{sec:data}

SpecPT is developed as a universal redshift measurement tool, designed to address the complexities that may arise from applying it to different datasets. The model's strength lies in its ability to generalize across diverse spectroscopic data by being trained on a sufficiently large and varied dataset. Similar to how large language models are trained on extensive corpora, SpecPT requires a vast and high-quality spectroscopic dataset to learn the inherent patterns of galaxy spectra. Once trained on such a dataset, the model can be fine-tuned to calibrate for other instruments with minimal additional data.

The DESI survey provides the ideal dataset for this purpose. Spanning five years, beginning in 2019, DESI aims to collect spectroscopic redshifts for over 35 million galaxies and quasars across 14,000 square degrees of the sky \citep{aghamousa2016desi}. The full dataset, is expected to cover nearly 80\% of the universe's history and will be invaluable for training a universal transformer model for spectroscopic data. For this paper, we leverage data from DESI's Early Data Release (EDR), specifically the One-Percent Survey \citep{adame2024validation}. This subset, covering roughly 1\% of the final DESI footprint, is well-suited for developing and testing the SpecPT model architecture and establishing an early proof of concept.

\subsection{Overview of the One-Percent Survey and DESI Programs} \label{sec:desi}

The One-Percent Survey was crucial for validating the scientific program for DESI, providing a representative sample of the full DESI target classes while achieving high completeness in fiber assignment and redshift estimation. This survey, which provides a representative sample of the full DESI target classes, was instrumental in assessing the efficiency of automated routines for data acquisition and in generating a highly complete dataset for redshift classification across all target classes.

DESI operates under a tiered approach, employing three distinct programs: the dark program, the bright program, and the backup program \citep{schlafly2023survey}. The dark program targets Luminous Red Galaxies (LRGs), Emission Line Galaxies (ELGs), and quasars across a redshift range of $0.4 < z < 3.5$, observed under optimal conditions. The bright program focuses on brighter galaxies and Milky Way stars, observed under suboptimal conditions, while the backup program targets even brighter Milky Way stars and is observed under the poorest conditions. This tiered strategy maximizes observational efficiency and minimizes systematic uncertainties.

The spectroscopic data in the DESI Early Data Release (EDR) are processed using the `fuji' version \citep{guy2023spectroscopic} of the DESI spectroscopic data reduction pipeline that applies the spectroperfectionism algorithm \citep{bolton2010spectro} for spectrum extraction, followed by corrections for fiber-to-fiber variations and subtraction of empirically derived sky models. The fluxes in the spectra are calibrated using stellar model fits to standard stars, and the calibrated spectra are co-added across multiple exposures to produce the final processed spectra. The DESI EDR also includes redshift measurements derived from the Redrock algorithm, which estimates redshifts by minimizing the $\chi^2$ between observed spectra and synthetic models constructed from Principal Component Analysis (PCA) templates. This sophisticated data processing and redshift determination process ensures the high quality and reliability of the spectroscopic data used in SpecPT.

The spectroscopic data in the DESI Early Data Release (EDR) undergo processing through the 'fuji' version of the DESI data reduction pipeline, as detailed by \cite{guy2023spectroscopic}. This pipeline employs the \textit{spectroperfectionism} algorithm, introduced by \cite{bolton2010spectro}, to extract spectra with precision. After extraction, corrections are applied to account for variations between fibers, and a sky model, derived from sky fibers, is subtracted to remove background contamination. The resulting spectra are then calibrated in flux by fitting them to stellar models, ensuring accuracy across different observations. These calibrated spectra are combined across multiple exposures to generate the final dataset. Additionally, the DESI EDR includes redshift estimates produced by the Redrock algorithm (S. J. Bailey et al., in preparation), which determines redshifts by minimizing the $\chi^2$ difference between the observed spectra and synthetic models created from Principal Component Analysis (PCA) templates. This meticulous processing and redshift calculation ensure that the spectroscopic data used in SpecPT are of the highest quality and reliability.

\subsection{The Bright Galaxy Survey (BGS)} \label{sec:bgs}

\begin{figure}[h]
    \centering
    \centerline{\includegraphics[width=\columnwidth]{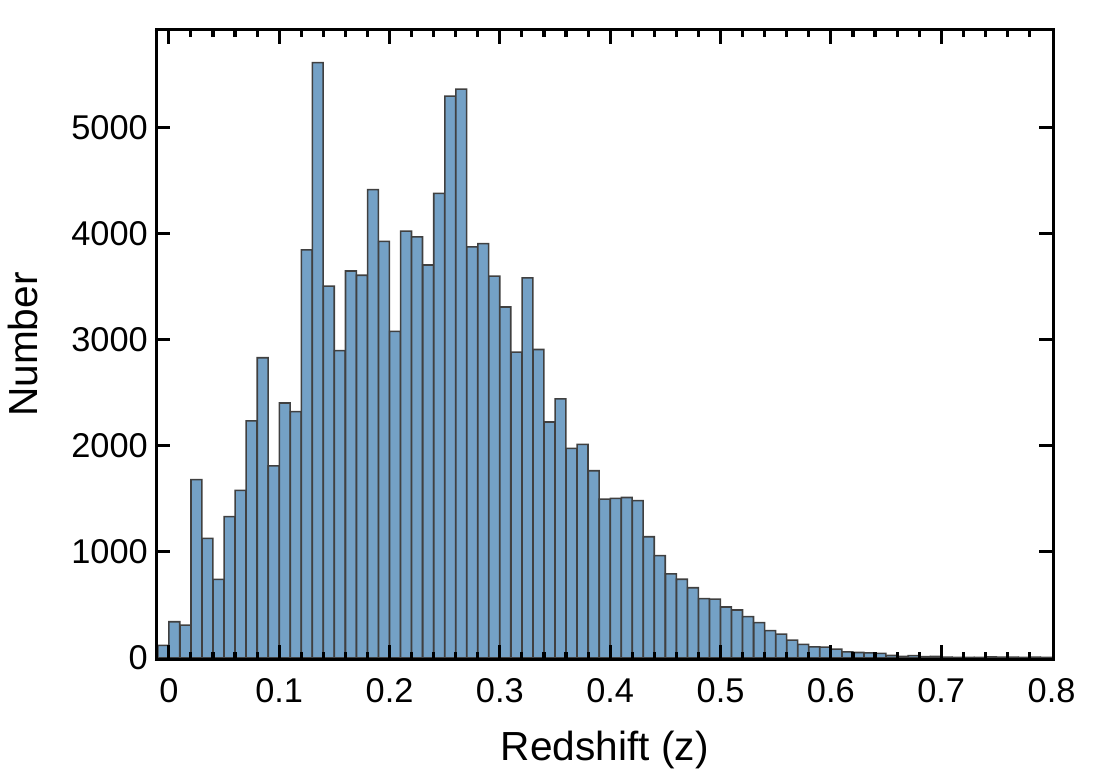}}
    \caption{This histogram shows the spectroscopic redshift distribution of the $\sim$129,000 objects in the BGS catalog. The peak of the distribution occurs around $z = 0.25$}
    \label{fig:bgs_zdist}
\end{figure}

The Bright Galaxy Survey (BGS, \citealt{hahn2023desi} is the cornerstone of DESI's efforts in low-redshift cosmology, designed to map the large-scale structure of the universe in the redshift range of $0.05 < z < 0.4$. BGS aims to ultimately observe over 10 million galaxies, with a strong emphasis on achieving high completeness and precise redshift measurements. The BGS sample is magnitude-limited, selected based on an r-band magnitude cutoff (BGS Bright) of $r \leq 19.5$ mag in DECaLS imaging areas, with slight adjustments in other regions to maintain uniform target density. A secondary, fainter subset (BGS Faint) extends to $r \leq 20.175$, allowing the exploration of less luminous galaxies.

To ensure the reliability of redshift measurements, \cite{hahn2023desi} suggest a series of stringent quality criteria. These criteria include:

\begin{enumerate}
    \item Selecting only spectra where no warning flags are raised by the Redrock algorithm and where the best-fit spectral type is classified as``galaxy."
    \item Ensuring the reported redshift error is sufficiently small, specifically $z_{err} < 0.0005 (1 + z)$, to maintain accurate redshift estimates.
    \item Requiring a high confidence level in the redshift measurement, defined by a difference of $\Delta\chi^2 > 40$ between the two best-fitting models, minimizing the risk of catastrophic redshift failures.
    \item Validating the reliability of the redshift against deep spectra, ensuring consistency within 1000 km s$^{-1}$: $|z_{\text{deep}} - z| / (1 + z_{\text{deep}}) < 0.0033$.
\end{enumerate}

These criteria are designed to filter out spectra with potential issues, such as spurious detections or significant uncertainties, thereby ensuring a redshift success rate of at least 95\% under typical observing conditions. The combination of these rigorous cuts allows the BGS to provide a high-quality, reliable dataset that serves as an excellent training set for SpecPT, particularly for low-redshift applications, where traditional spectroscopic methods usually excel.

After applying these cuts, the final BGS sample we use consists of approximately 129,000 spectra, as illustrated in Figure 2. The redshift distribution of this sample peaks at $z = 0.25$, providing a representative dataset for training and testing the model in the low-redshift regime.

\subsection{Emission Line Galaxy (ELG) Sample} \label{sec:elg}

\begin{figure}[h]
    \centering
    \centerline{\includegraphics[width=\columnwidth]{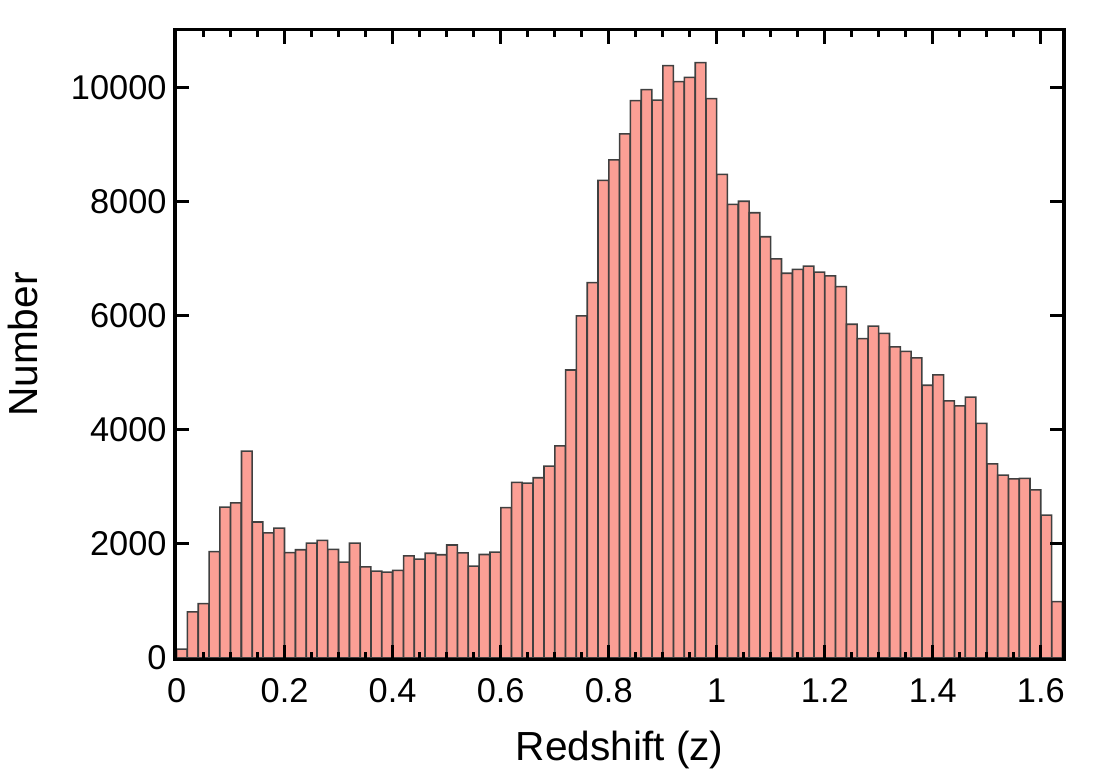}}
    \caption{This histogram shows the spectroscopic redshift distribution of the 371,671 Emission Line Galaxies (ELGs) in our dataset. The peak of the distribution occurs around $z = 0.85$}
    \label{fig:elg_zdist}
\end{figure}

The ELG sample is a key component of DESI’s intermediate to high-redshift cosmology, targeting star-forming galaxies in the $0.6 < z < 1.6$ range. This sample is expected to provide one-third of all DESI redshifts, making it a crucial dataset for training SpecPT across a broad redshift range. ELGs are selected based on their photometric properties, particularly in the g, r, and z bands, using a combination of magnitude cuts and color-space selections designed to isolate star-forming galaxies with strong [O II] emission.

The ELG targets are carefully chosen to maximize redshift success rates while minimizing contamination from low-redshift interlopers and stars. The selection strategy also prioritizes ELGs in the $1.1 < z < 1.6$ range, where DESI anticipates deriving its most stringent cosmological constraints. The DESI spectrographs are optimized to resolve key emission features, such as the [O II] doublet, which is crucial for the ELG sample. 

An important aspect of constructing a reliable ELG sample involves applying stringent quality cuts to ensure the accuracy of redshift measurements. We use the criteria established by \cite{raichoor2023target} to select spectra with reliable redshift measurements. This involves applying a combination of cuts based on the difference in chi-squared values ($\Delta\chi^2$) between the best and second-best redshift fits and the signal-to-noise ratio of the [O II] emission line ($\texttt{FOII\_SNR}$). A high $\Delta\chi^2$ value typically indicates a more secure redshift, while the $\texttt{FOII\_SNR}$ helps to account for cases where a high-confidence redshift may still be obtained from a low signal-to-noise spectrum dominated by the [O II] doublet. The dual criteria are designed to maximize the inclusion of accurate redshifts while minimizing the incidence of catastrophic errors, effectively balancing completeness and reliability in the final ELG sample.

After implementing the quality cuts to ensure reliable redshift measurements, our final sample comprises 371,671 ELG spectra, which are used for training and testing SpecPT. The redshift distribution of this ELG sample is illustrated in Figure \ref{fig:elg_zdist}, where we observe that the distribution peaks around $z = 0.85$.

The ELG sample serves as an essential intermediate redshift training set for SpecPT, with its focus on galaxies featuring strong emission lines, providing an ideal dataset for developing automated techniques that can outperform traditional methods in speed and efficiency.

\subsection{Data Preparation for SpecPT} \label{sec:data_prep}

To prepare the data for SpecPT, additional pre-processing steps are applied to the spectroscopic data from the BGS and ELG samples. Spectra are normalized by dividing them by their median flux values and resampled to a common wavelength grid, ensuring consistency across different redshift ranges. These pre-processing steps are necessary for preparing the data to be effectively used in training a universal redshift measurement model. We develop a SpecPT model for each of the two datasets: a low-z model for the BGS sample and a high-z model for the ELG sample.

\section{SpecPT Architecture}
\label{sec:SpecPT}

\begin{figure*}
    \includegraphics[width=\textwidth]{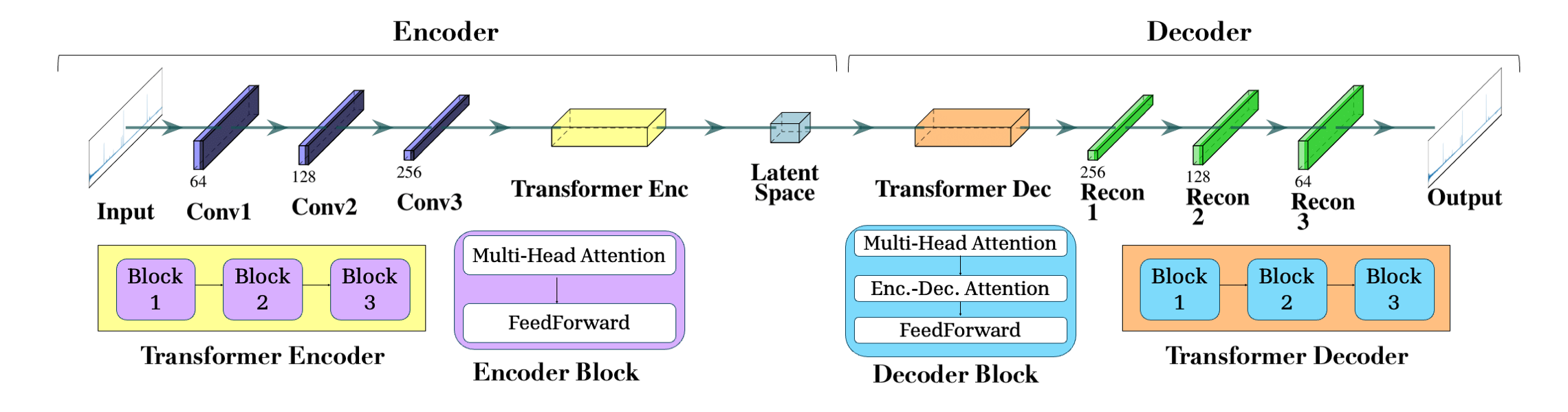}
    \caption{The SpecPT autoencoder architecture begins with the observed spectrum input, processed through convolutional blocks (comprising convolutional and max pooling layers) for feature extraction, followed by a transformer encoder generating compact embeddings. The decoder, featuring a transformer decoder and linear reconstruction layers, then meticulously reconstructs the original observed spectrum.}
    \label{fig:specpt_main}
\end{figure*}

The SpecPT architecture introduces an advanced autoencoder framework, specifically designed for analyzing spectroscopic data. This architecture utilizes a combination of convolutional layers for initial feature extraction, followed by transformer encoders and decoders to handle complex patterns in the data. The entire architecture of the SpecPT model (shown in Figure \ref{fig:specpt_main}) is designed to capture and compress the most important features of spectroscopic data, while the SpecPT for redshift model (illustrated in Figure \ref{fig:specpt_redshift}) builds upon this framework with additional modifications tailored for redshift estimation. This section details each component of the SpecPT model, explaining its function and role in the overall architecture.

\subsection{Encoder} \label{sec:encoder}

The encoder is an essential part of the SpecPT architecture, responsible for extracting key spectral features from the input data while compressing them into a compact representation that retains the necessary information for subsequent tasks.

\subsubsection{Convolutional Feature Extraction}

The encoder begins with a series of one-dimensional convolutional layers. These layers progressively reduce the dimensionality of the input spectrum, while simultaneously extracting important features such as emission lines, absorption lines, and continuum shapes. Specifically, three 1D convolutional layers are used with kernel sizes of 41, 21, and 11, respectively, followed by batch normalization and ReLU activation functions. This choice of kernel sizes allows for the detection of features at different scales, with larger kernels better suited for broader spectral features and smaller kernels for fine details.

The convolutional layers are followed by max pooling, which further reduces the dimensionality while retaining the most relevant information. This design is inspired by similar approaches used in other spectroscopic analysis models, which demonstrated that convolutional layers are effective for capturing localized spectral features, ensuring that the model can accurately capture the various components of a galaxy spectrum (see, for e.g., \cite{melchior2023autoencoding,zhong2023galaxy,stivaktakis2019convolutional,zhou2021spectroscopic,wu2020predicting,fabbro2018application}). Max pooling, in particular, helps to make the model more robust by focusing on the most prominent spectral features while discarding less relevant noise.

\subsubsection{Transformer Encoder}

Following the convolutional feature extraction, the model uses a Transformer Encoder to process the compressed spectral data. The Transformer Encoder consists of three encoder layers, each with eight attention heads. The attention mechanism is key to capturing long-range dependencies within the spectral data, which are often critical for understanding the physical properties of galaxies. For instance, relationships between emission and absorption lines that are far apart in wavelength space can provide insights into stellar populations, ionization states, and star formation rates (see, for e.g., \cite{melchior2023autoencoding,liang2023outlier,liang2023autoencoding}).

The Transformer Encoder uses self-attention to dynamically focus on the most relevant portions of the spectrum for a given task, allowing it to capture complex interactions between different spectral features. This ability to capture both local and global patterns makes Transformer-based models highly effective for analyzing spectroscopic data, where such patterns are often indicative of underlying astrophysical processes. The use of feedforward layers following each attention layer helps to further process the attention outputs, adding non-linearity and allowing the model to learn more complex feature representations.

\begin{figure*}
    \includegraphics[width=\textwidth]{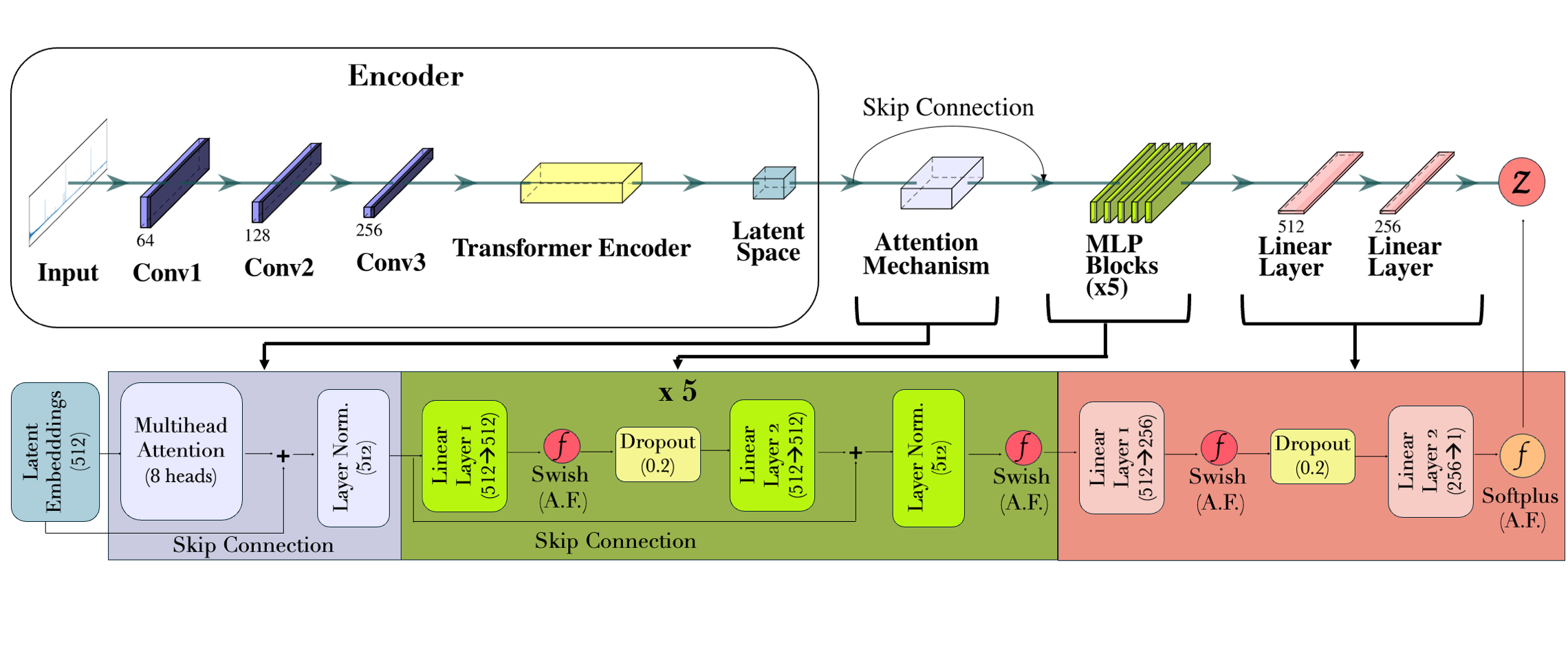}
    \caption{The SpecPT redshift prediction architecture integrates the fully trained SpecPT encoder with residual MLP blocks, an attention mechanism, and linear output layers to transform embeddings into redshift predictions.}
    \label{fig:specpt_redshift}
\end{figure*}

\subsection{Decoder} \label{sec:decoder}

The decoder reconstructs the input spectrum from the compressed representation generated by the encoder. It uses a transformer decoder, which mirrors the structure of the transformer encoder. The decoder layers apply cross-attention, which helps align the latent representation generated by the encoder with the original spectral features. This cross-attention allows the decoder to focus on specific parts of the latent representation that are most relevant for reconstructing the spectrum.

The reconstruction is completed by passing the output through a series of fully connected layers. These linear reconstruction layers progressively transform the latent representation back to the original input size, ensuring that the reconstructed spectrum matches the input spectrum as closely as possible. The goal of the decoder is to minimize reconstruction loss, ensuring that the encoded latent features contain all the necessary information to accurately represent the input spectrum.

\subsection{SpecPT for Redshift Prediction} \label{sec:mlp}

The SpecPT for redshift model, illustrated in Figure \ref{fig:specpt_redshift}, builds upon the encoder-decoder framework of SpecPT but incorporates additional layers specifically designed for redshift prediction. These modifications include residual Multi-Layer Perceptron (MLP) blocks, attention mechanisms, and a specialized output layer to ensure accurate redshift estimation.

\subsubsection{Residual MLP Blocks}

The redshift prediction module includes a series of five Residual MLP Blocks. These blocks consist of fully connected layers with skip connections, which help mitigate the issue of vanishing gradients and allow the model to learn more effectively by preserving information from earlier layers. Each residual block uses Swish activation functions, which provide smoother gradients and help improve learning compared to traditional ReLU activations. Swish activation, represented by $f(x)$, is defined as,

\begin{equation}
    f(x) = x \times \text{sigmoid}(x) = x \times \frac{1}{1 + e^{-x}}\mperiod
\end{equation}

This is particularly important for spectroscopic data, where subtle differences in feature intensities can have significant implications for the predicted redshift.

The residual connections in the MLP blocks are inspired by their success in deep learning models used in computer vision \citep{he2016deep} and natural language processing \citep{conneau2016very}. They allow for deeper networks without performance degradation, ensuring that the model can learn complex, hierarchical features that are necessary for precise redshift prediction.

\subsubsection{Attention Mechanism}

In addition to the residual MLP blocks, an attention mechanism is used to further refine the latent representation before redshift prediction. Specifically, a multi-head attention layer with eight heads is applied to the encoded features. This attention mechanism helps the model weigh different parts of the latent representation, focusing more on those features that are most informative for determining the redshift. By using residual connections around the attention layer, the model can retain the original encoded features while incorporating the additional information provided by the attention mechanism.

\subsubsection{Redshift Output Layer}

The Redshift Output Layer consists of a linear layer followed by a Softplus activation function. Softplus activation function is defined as follows,
\begin{equation}
    g(x) = \ln(1 + e^x)\mperiod
\end{equation}
The linear layer transforms the refined latent representation into a single scalar value, which represents the predicted redshift. The Softplus activation ensures that the output is strictly positive, aligning with the physical constraints of redshift values. This activation function also helps prevent numerical instabilities that can arise from negative predictions, making the model more robust during training.

\begin{figure*}
    \includegraphics[width=\textwidth]{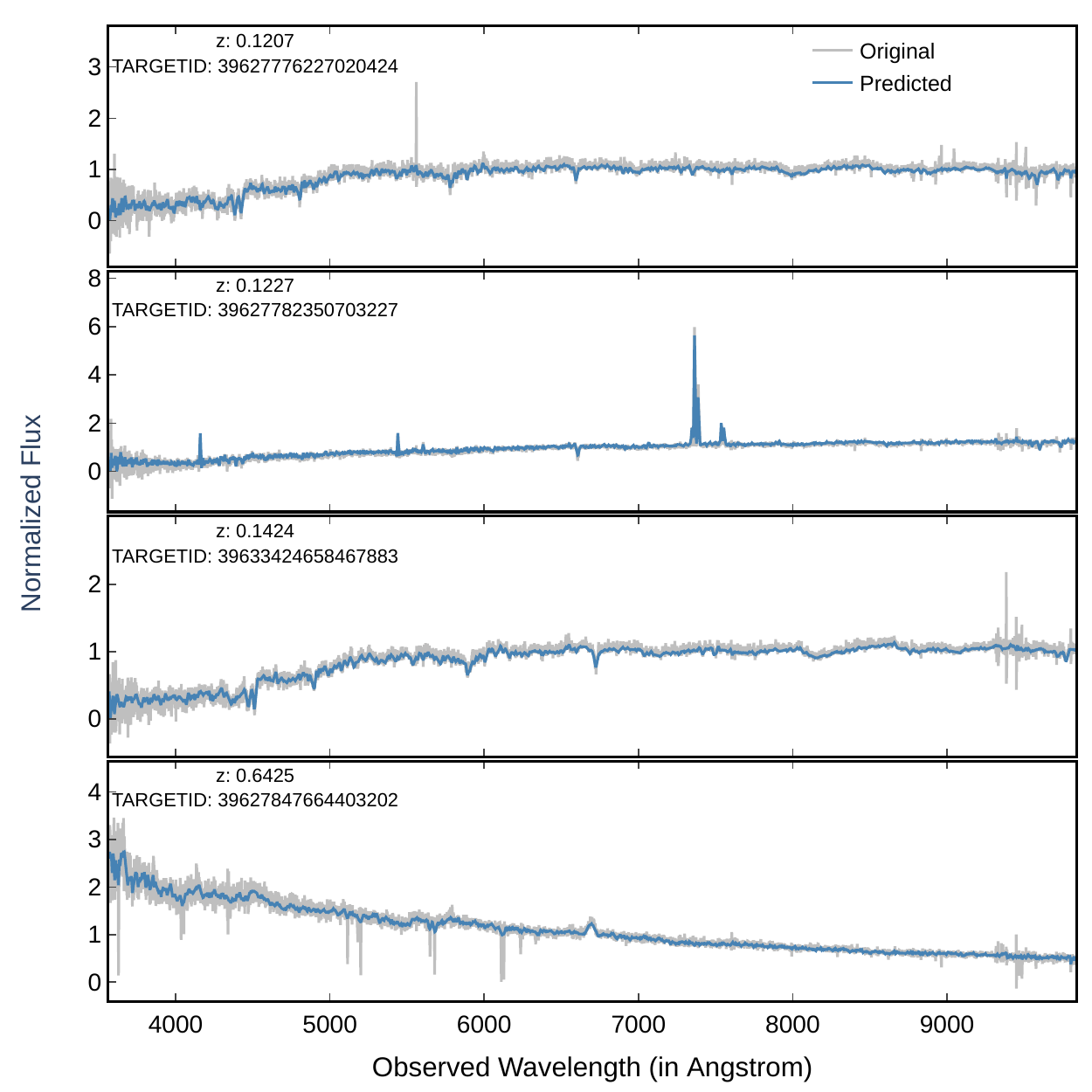}
    \caption{Examples of original and reconstructed spectra from the BGS dataset, presented in the observed frame with increasing redshift from top to bottom. Each panel shows the original spectrum (gray) alongside the autoencoder reconstruction (blue), demonstrating strong alignment between the two. The autoencoder accurately reconstructs key spectral features, such as emission and absorption lines, while effectively reducing noise levels.}
    \label{fig:bgs_auto}
\end{figure*}

\begin{figure*}
    \includegraphics[width=\textwidth]{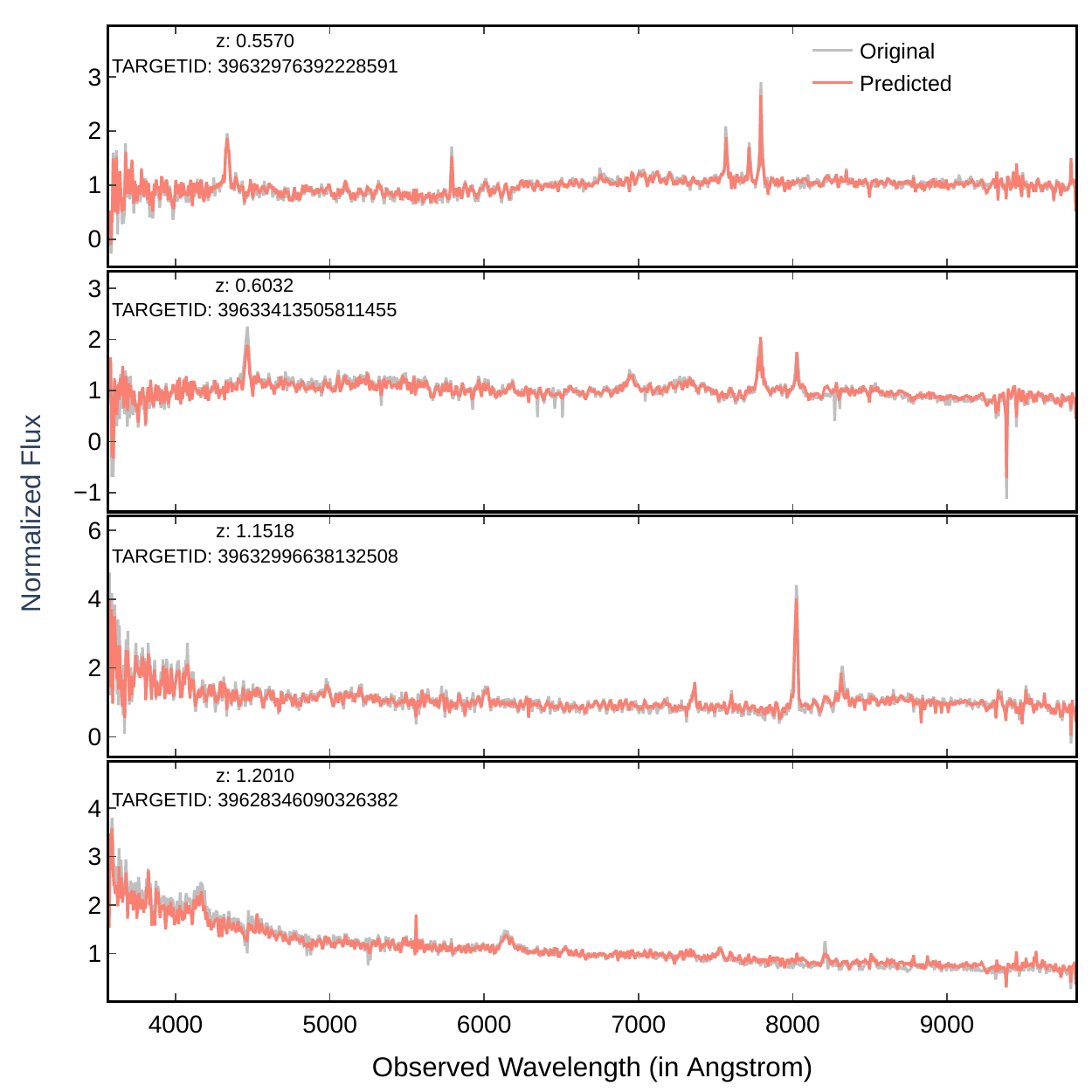}
    \caption{Examples of original and reconstructed spectra from the ELG dataset, shown in the observed frame with increasing redshift from top to bottom. Each panel compares the original spectrum (gray) with the autoencoder reconstruction (red), highlighting the model's ability to capture essential features like emission lines and continuum shape. Similar to the BGS results, the ELG reconstructions reveal reduced noise, allowing for precise recovery of spectral details critical for redshift estimation.}
    \label{fig:elg_auto}
\end{figure*}

\subsection{Training and Implementation} \label{sec:training}

The SpecPT model is trained using a novel loss function, the \textit{Normalized \textbf{Mean} Absolute Deviation} (\nmad) loss, which we developed specifically for this study. The function \nmad is inspired by the \textit{Normalized \textbf{Median} Absolute Deviation (NMAD)} metric \citep{hoaglin2000understanding}, commonly used to measure catastrophic photometric redshift outliers \citep{ilbert2008cosmos}. Mathematically, the \nmad offers a robust alternative to traditional loss functions like Mean Squared Error (MSE) or Mean Absolute Error (MAE) by reducing the impact of outliers, leading to improved convergence and more accurate predictions for spectroscopic data. The \nmad is defined as:

\begin{equation}
\text{\nmad} = \frac{\text{mean}(|z_{\text{pred},i} - z_{\text{true},i}|)}{\text{std}(z_{\text{true}})}\mcomma
\end{equation}

\noindent where $z_{pred,i}$ represents the predicted redshift values, $z_{true,i}$ represents the true redshift values, $\text{mean}(|z_{\text{pred},i} - z_{\text{true},i}|)$ is the Mean Absolute Deviation (MAD) of the predicted and true redshifts, and $\text{std}(z_{true)}$ is the standard deviation of the true redshift values. This formulation enhances the model's ability to learn from challenging and noisy datasets, as demonstrated by our training.

The SpecPT autoencoder model, comprising 1,120,475,621 trainable parameters, requires approximately 43 hours to train, while the SpecPT model for redshift prediction, with 74,016,257 parameters, takes around 6 hours on a NVIDIA Quadro RTX 8000 GPU. Both models are implemented using the \textsc{pytorch} library \citep{paszke2019pytorch}, a powerful deep learning library that facilitates efficient model training and deployment.

To validate the training process, we employed the k-Fold cross-validation technique, where the test set remained fixed while the training and validation splits varied for each fold. Specifically, the dataset excluding the test set was split into 10 equal parts, with one part used as the validation set and the remaining nine as the training set. This process was repeated 10 times, with a different fold used as the validation set each time and allows us to test the robustness of our model.

\section{Results}
\label{sec:results}

\begin{figure}
    \centering
    \includegraphics[width=\columnwidth]{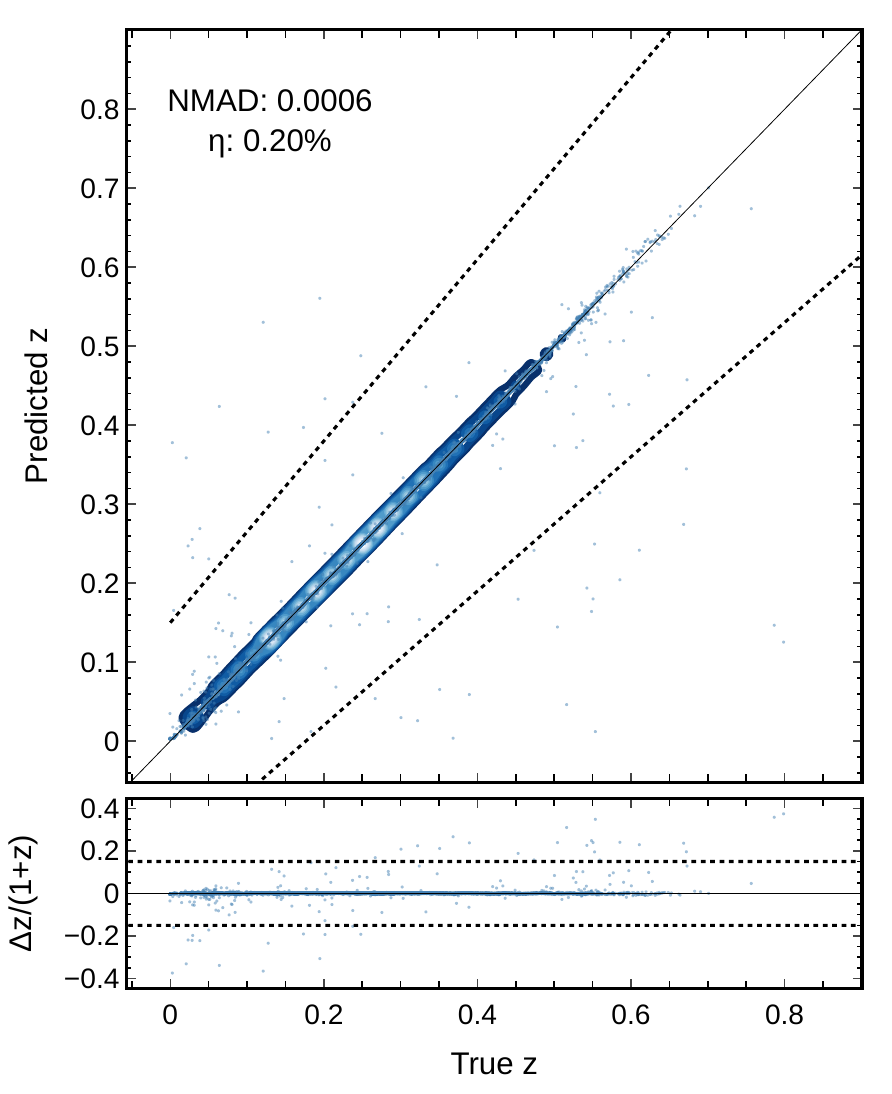}
    \caption{{SpecPT redshift measurement results for 19,351 objects in the BGS test set. The top panel shows predicted vs.\ true redshifts with density contours, closely aligning with the $y=x$ line. The bottom panel presents the distribution of normalized redshift residuals, $\Delta z/(1+z)$, centered near zero. Dotted lines mark the catastrophic outlier threshold ($|\Delta z|/(1+z) > 0.15$). The low NMAD and minimal outlier fraction ($\eta$) highlight SpecPT’s high precision and reliability.}}
    \label{fig:bgs_z_density}
\end{figure}

\begin{figure}
    \centering
    \includegraphics[width=\columnwidth]{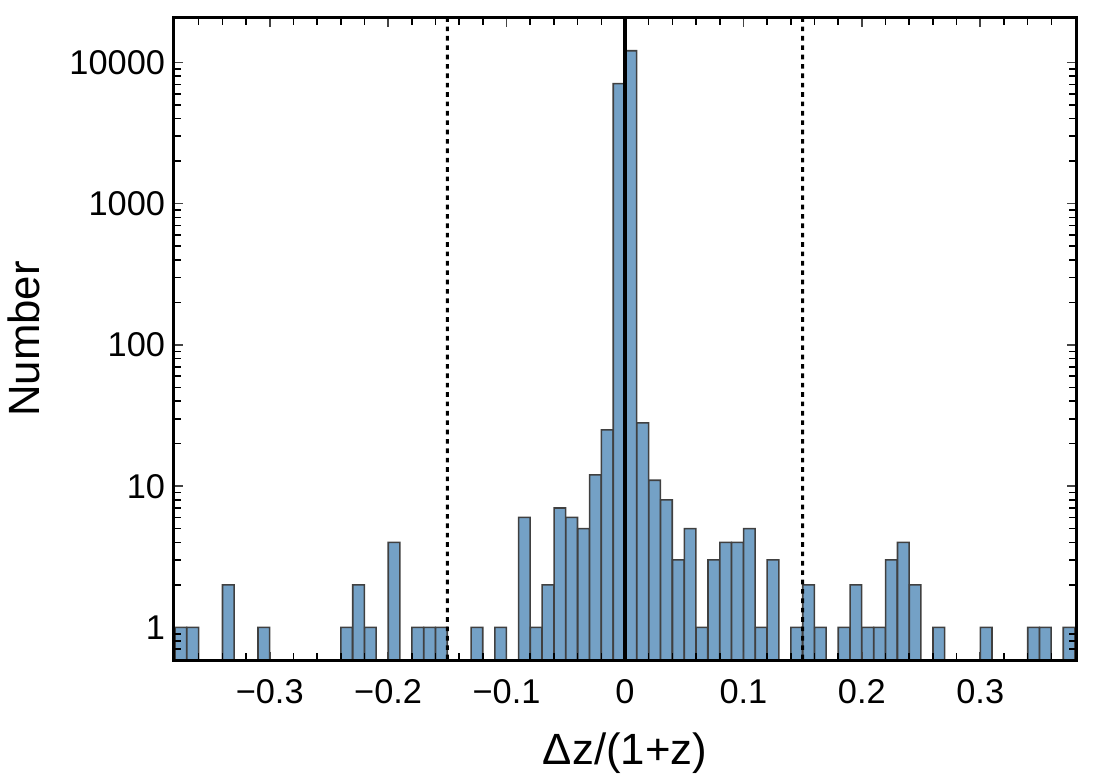}
    \caption{{Histogram of normalized redshift residuals, $\Delta z/(1+z)$, for the BGS dataset. The distribution peaks at 0, indicating highly accurate redshift predictions with minimal deviation. Very few objects exceed the catastrophic outlier threshold ($|\Delta z|/(1+z) > 0.15$, dotted lines), illustrating SpecPT’s exceptional performance and reliability.}}
    \label{fig:bgs_z_error}
\end{figure}

In this section, we present the performance results of the SpecPT model on the BGS and ELG datasets. Our evaluation covers both the autoencoder’s ability to reconstruct spectra and the accuracy of the redshift predictions. For assessing redshift prediction, we employ NMAD, defined as
\begin{equation}
\text{NMAD} = 1.48 \times \text{median}\left(\left|\frac{\Delta z}{1+z_{\text{true}}}\right|\right)\mcomma
\end{equation}
where $\Delta z = (z_{\text{true}} - z_{\text{predicted}})$, following the metric used by \cite{ilbert2008cosmos} to compare the quality of photometric and spectroscopic redshifts. NMAD is directly comparable to the rms/(1+$z$) metric commonly used for evaluating redshift accuracy. In addition, we apply a criterion to identify catastrophic outliers, defined as $\left|\Delta z\right| / (1+z) > 0.15$, also following \cite{ilbert2008cosmos}. The fraction of outliers in the dataset, denoted by $\eta$, measures the model’s reliability in predicting accurate redshifts. We also present visualizations of the embedding space to illustrate how the model naturally differentiates spectra based on redshift, and we analyze redshift prediction uncertainties (obtained using K-Fold training process as described in Section \ref{sec:training}) across different redshift and z-band magnitude bins to further evaluate the model’s performance.

\subsection{Autoencoder reconstruction} \label{sec:auto_results}

The first step in developing the SpecPT model involves training an autoencoder to learn efficient latent representations of galaxy spectra. The autoencoder was trained separately for the BGS and ELG datasets to minimize the reconstruction loss between the original input spectra and their reconstructions after passing through the encoder-decoder network. This process allows the model to capture the essential spectral features while discarding irrelevant noise.

In Figures \ref{fig:bgs_auto} and \ref{fig:elg_auto}, we present four examples each from the BGS and ELG datasets, showing a comparison between the original spectra and their reconstructions. The reconstructed spectra capture key features with remarkable accuracy, including prominent emission and absorption lines, as well as the general shape of the continuum. The ability of the model to reproduce these features is crucial, as they contain the information needed for the subsequent redshift prediction.

An interesting observation is that the autoencoder appears to differentiate between genuine spectral features and noise artifacts. For instance, in the case of TARGETID = 39627776227020424 from the BGS sample (see Figure \ref{fig:bgs_auto}), the model avoids reconstructing the skyline emission artifact observed around 5560 \r{A}, indicating that it effectively reduces sky contamination while retaining critical spectral information. This behavior suggests that the model is learning to recognize and filter out noise, thereby improving the quality of the spectral data.

Overall, the reconstructed spectra show a general reduction in noise levels, while maintaining all important features, such as emission and absorption lines. These results demonstrate the success of the autoencoder in capturing meaningful latent representations, which are essential for enhancing the accuracy of the downstream redshift prediction task.

\subsection{Redshift measurement} \label{sec:redshift_results}

After training the autoencoder, we adapt the trained encoder by integrating additional layers, as illustrated in Figure \ref{fig:specpt_redshift}, to construct a model capable of predicting redshifts. As discussed in Section \ref{sec:SpecPT}, the primary approach involves initially training the model to capture the inherent features within the spectra. This pre-training step is essential for enabling the model to learn detailed representations of spectral data, which can then be further optimized to perform specific tasks, such as redshift prediction, with minimal bias towards a single task. This strategy is particularly beneficial for SpecPT, as it is intended to function as a foundational model for spectroscopic analysis. In this section, we present the results of SpecPT’s performance on the test data for redshift measurement, focusing on each of the two primary datasets: BGS and ELG samples.

\begin{figure*}
    \includegraphics[width=\textwidth]{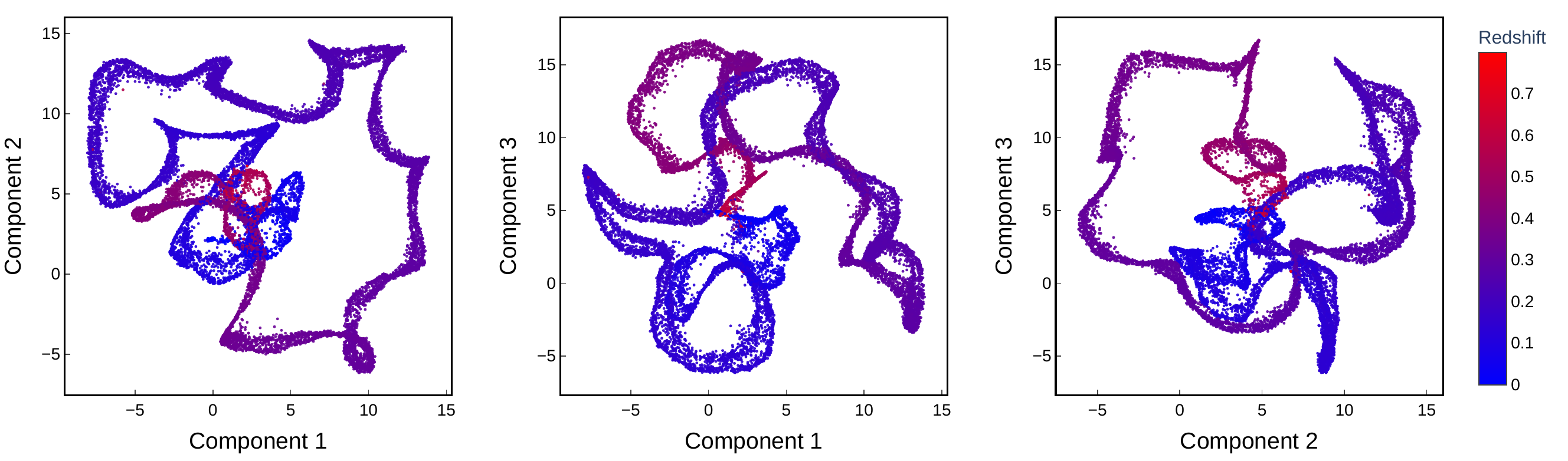}
    \caption{{UMAP visualization of the latent space embeddings from the SpecPT model for the BGS dataset. Each panel shows pairwise plots of the three UMAP components, with points colored by redshift (blue to red for increasing redshift). The smooth gradient reflects the model's ability to group spectra by redshift, highlighting its effective feature learning.}}
    \label{fig:bgs_umap}
\end{figure*}

\begin{figure*}
    \includegraphics[width=\textwidth]{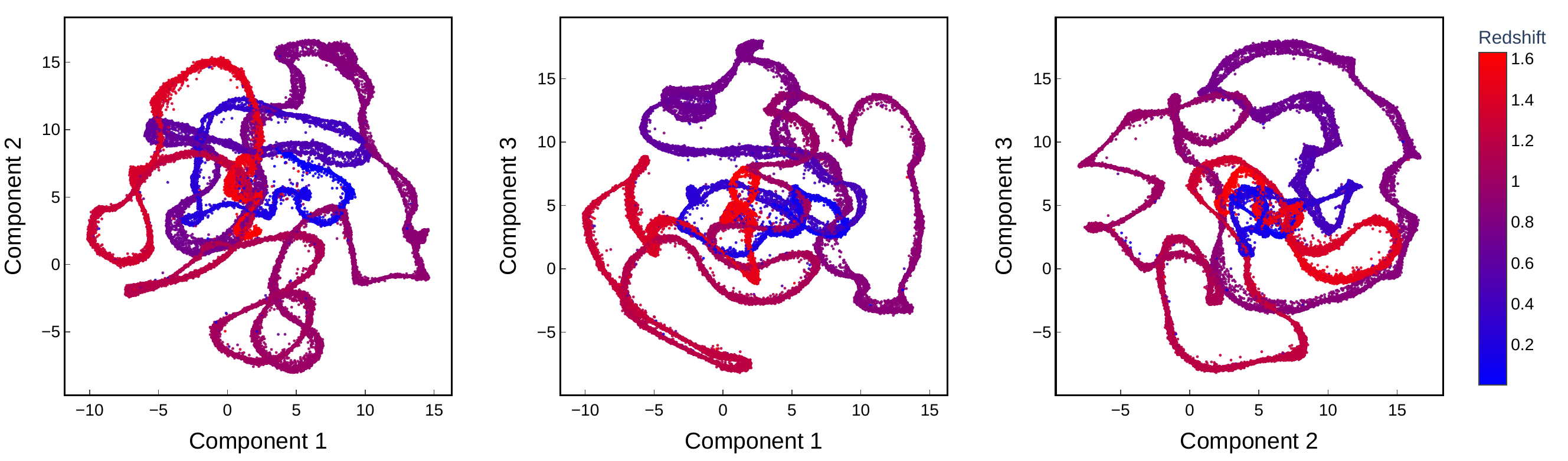}
    \caption{{UMAP visualization of ELG latent space embeddings, colored by redshift. Pairwise plots of the three components show a clear gradient from blue to red, reflecting increasing redshift. This highlights SpecPT’s ability to effectively group spectra by redshift, demonstrating robust performance for higher-redshift ELG data.}}
    \label{fig:elg_umap}
\end{figure*}

\subsubsection{BGS data} \label{sec:redshift_results_bgs}

The BGS dataset consists of 129,024 spectra. For the purpose of training and evaluation, this dataset was divided into three subsets: 70\% for training, 15\% for validation, and 15\% for testing. The results presented here are based on the test set, comprising 19,354 spectra.

To assess the quality of predictions, we compare the true redshifts to the predicted redshifts in Figure \ref{fig:bgs_z_density}. The top panel shows a scatter plot with density contours of predicted versus true redshifts, where the strong alignment with the $y=x$ line reflects the accuracy of the model's predictions. The bottom panel visualizes the normalized residuals, $\Delta z/(1+z)$, where the majority of data points cluster tightly around $\Delta z/(1+z) = 0$, further emphasizing the precision of SpecPT. This analysis results in an NMAD value of 0.0006, indicating a high degree of accuracy in the predictions. Moreover, the catastrophic outlier fraction, $\eta$, is just 0.20\%, showcasing the model's reliability in accurately predicting redshifts for the vast majority of objects.

Figure \ref{fig:bgs_z_error} provides a deeper look at the distribution of normalized residuals. The histogram reveals that the residuals are centered sharply at zero, with the majority of data points lying well within the catastrophic outlier threshold ($|\Delta z|/(1+z) > 0.15$), marked by vertical dotted lines. The logarithmic y-axis highlights that outliers are exceedingly rare, numbering in the tens within a dataset of around 19,000 spectra. This further reinforces the robustness of SpecPT for redshift prediction, even at the tails of the distribution.

Additionally, we analyze the latent space embeddings generated by the SpecPT model to understand how the spectra are represented internally. Using UMAP to reduce the latent space to three components, we visualize the embeddings in Figure \ref{fig:bgs_umap}. The three pairwise scatter plots of the components, colored by redshift, show a smooth and continuous gradient of redshift values. This clear structure in the latent space demonstrates that SpecPT effectively groups spectra by their redshift, capturing the inherent relationships between spectral features and redshift in a way that facilitates accurate predictions.

\subsubsection{ELG data} \label{sec:redshift_results_elg}

\begin{figure}
    \centering
    \includegraphics[width=\columnwidth]{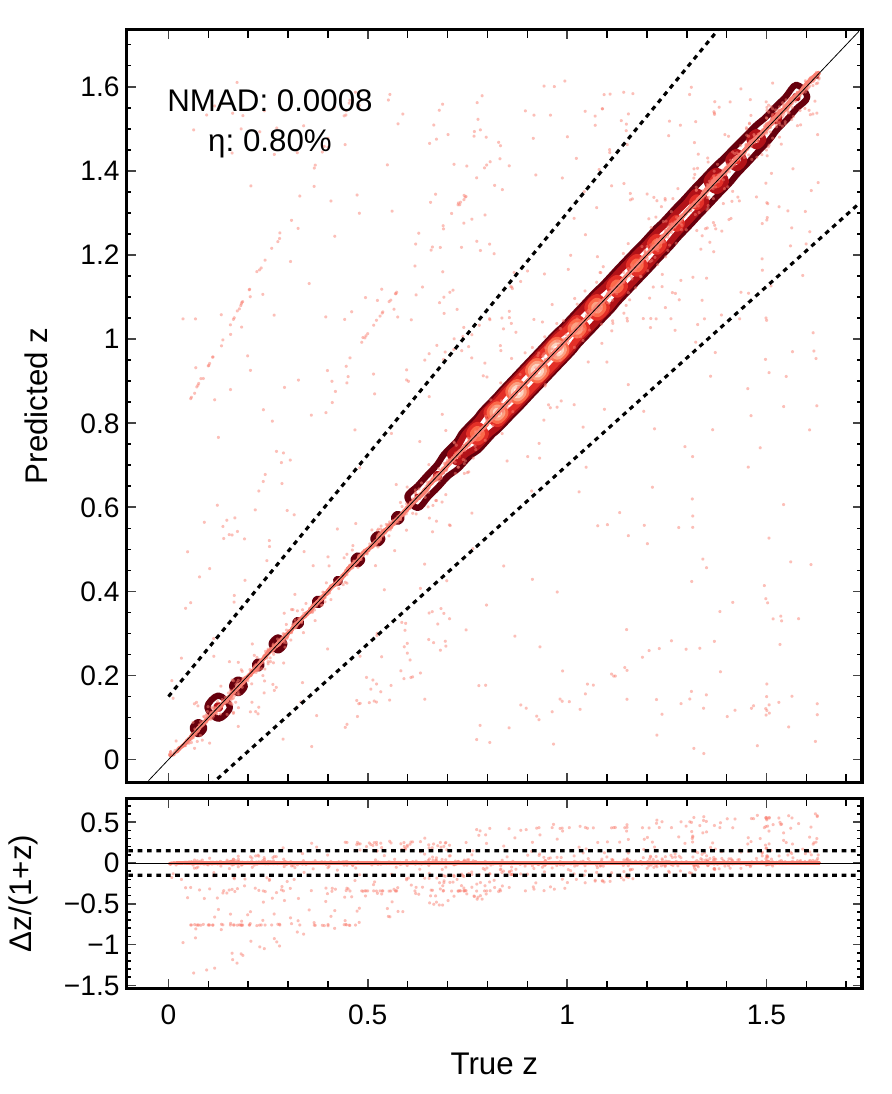}
    \caption{{SpecPT redshift prediction results for 55,741 ELG test set objects. The top panel shows predicted versus true redshifts with density contours, closely lying along the $y=x$ line, indicating high accuracy. The bottom panel displays normalized redshift residuals, $\Delta z/(1+z)$, clustering around 0, with few catastrophic outliers ($|\Delta z|/(1+z) > 0.15$, dotted lines). Low NMAD and outlier fraction ($\eta$) confirm robust and precise performance, consistent with BGS results (Figure \ref{fig:bgs_z_density}).}}
    \label{fig:elg_z_density}
\end{figure}

\begin{figure}
    \centering
    \includegraphics[width=\columnwidth]{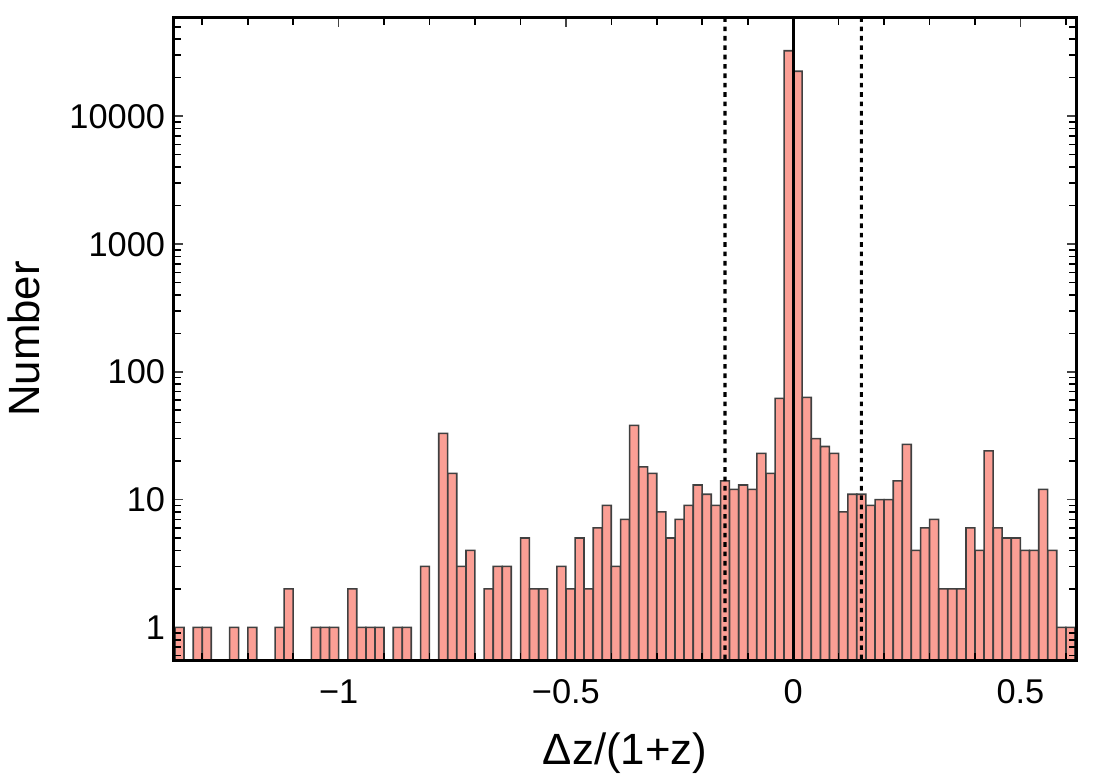}
    \caption{{Histogram of normalized redshift residuals, $\Delta z/(1+z)$, for the ELG dataset. The distribution peaks at 0, indicating minimal deviations in most predictions, with very few objects exceeding the catastrophic outlier threshold ($|\Delta z|/(1+z) > 0.15$, dotted lines). These results, consistent with BGS trends (Figure \ref{fig:bgs_z_error}), highlight SpecPT’s accuracy and reliability in redshift estimation.}}
    \label{fig:elg_z_error}
\end{figure}

For the ELG dataset, SpecPT demonstrates similarly robust performance as observed with the BGS dataset, even in the higher redshift range of $0.6 < z < 1.6$. For the 371,671 spectra in the ELG dataset, we reserved 15\% (55,741 spectra) as the test set, with results from this subset presented here.

Figure \ref{fig:elg_z_density} shows the predicted versus true redshift values, where most data points closely follow the $y=x$ line, as illustrated by the density contours. The normalized redshift residuals, $\Delta z/(1+z)$, in the bottom panel of the same figure, similarly cluster around the $\Delta z/(1+z) = 0$ line, emphasizing the model’s accuracy. The NMAD for this dataset is $0.0008$, slightly higher than the $0.0006$ value observed for the BGS dataset, but still within the same order of magnitude, indicating consistently strong performance. The outlier fraction, $\eta = 0.80\%$, though higher than the BGS fraction, remains impressively low for a sample with inherently noisiZSXer, more complex spectra and fainter objects.

The histogram of normalized residuals in Figure \ref{fig:elg_z_error} further reinforces these observations. The residuals form a sharp peak at 0, with the y-axis (log scale) highlighting how few objects lie beyond the catastrophic outlier threshold of $|\Delta z|/(1+z) > 0.15$. This minimal scatter around the central peak indicates SpecPT’s precision in predicting redshifts across the ELG dataset, even for challenging high-redshift sources.

The latent space embeddings, visualized using UMAP in Figure \ref{fig:elg_umap}, provide additional validation of the model’s reliability. Similar to the results for the BGS dataset, the embeddings exhibit a smooth gradient in redshift, transitioning from low (blue) to high (red). This clear organization demonstrates the model’s ability to differentiate spectral features corresponding to redshift, further supporting the notion that the latent representations learned by SpecPT are both meaningful and generalizable.

\subsection{Redshift residuals as a function of Redshift} \label{sec:residual_results}

\begin{figure}
    \centering
    \includegraphics[width=\columnwidth]{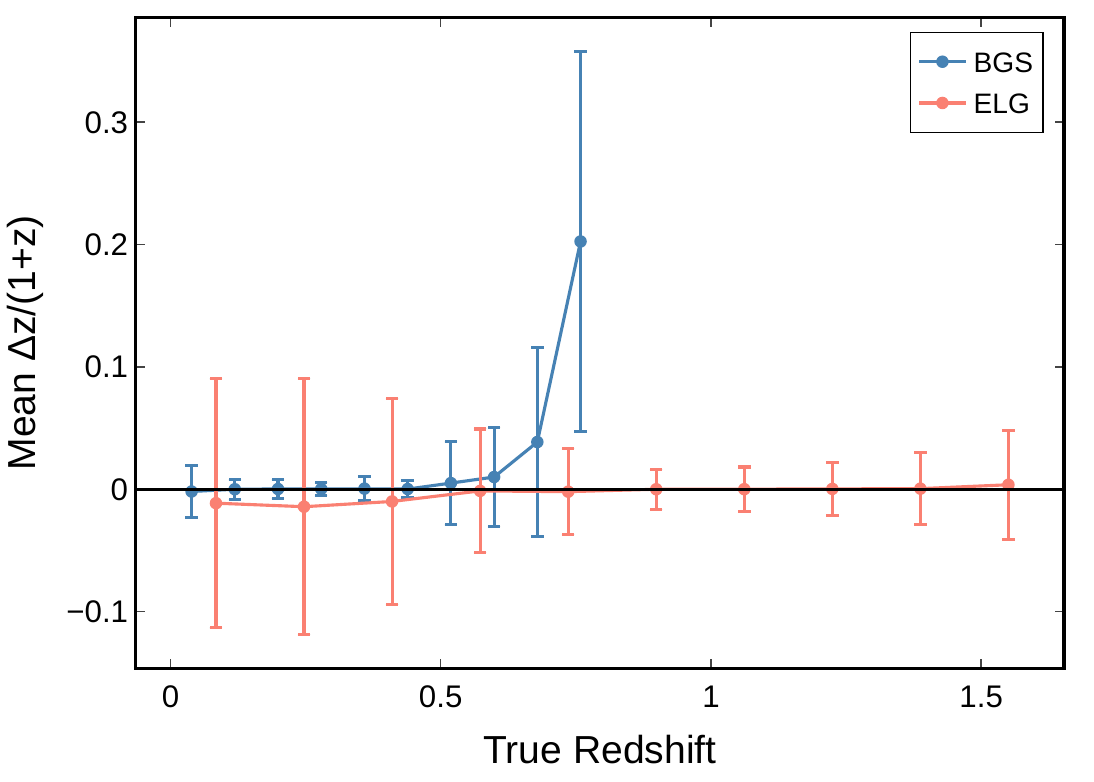}
    \caption{{Mean redshift residuals as a function of redshift for BGS (blue) and ELG (red) datasets. Points represent mean residuals in 10 redshift bins, with error bars denoting standard deviations. The trend reflects the data distribution of each dataset, with lower residuals and variability in well-represented redshift ranges and increased errors in sparsely populated regions.}}
    \label{fig:zbin_error}
\end{figure}

To further evaluate SpecPT's performance across different redshift ranges, we plot the mean residuals, calculated as $\Delta z/(1+z)$, for 10 redshift bins for both the BGS (blue) and ELG (red) datasets in Figure \ref{fig:zbin_error}. The markers indicate the mean residuals in each bin, while the error bars represent the standard deviation of the residuals within the respective bins. These plots provide a detailed view of how the redshift prediction errors vary with redshift across the two datasets.

For the BGS dataset, the standard deviation of the residuals initially decrease with increasing redshift, showing both lower mean values and reduced standard deviations. However, at higher redshifts, there is a noticeable increase in both the mean residuals and the standard deviation. This trend aligns closely with the redshift distribution of the BGS dataset, as shown in Figure \ref{fig:bgs_zdist}. The larger errors at higher redshifts correlate with the reduced amount of training data available in this range, while the best performance occurs in the redshift range with the densest data coverage.

A similar trend is observed for the ELG dataset. The standard deviation of residuals decrease with increasing redshift, reflecting the higher density of ELG data at higher redshifts, as illustrated in Figure \ref{fig:elg_zdist}. At the lower redshift end, where data is sparse, the model exhibits higher standard deviation of residuals and variability.

An important observation is that the performance of SpecPT across the two datasets appears complementary. For the BGS dataset, SpecPT achieves its best results at low redshifts, with performance degrading at higher redshifts. Conversely, for the ELG dataset, the model performs well at higher redshifts, with more variability at lower redshifts. This complementary behavior can be attributed to the differences in the redshift distributions of the two datasets, with BGS dominating at lower redshifts and ELG providing more data at higher redshifts.

\section{Discussion}
\label{sec:discussion}

In this section, we explore the implications of the results presented in this paper, focusing on SpecPT’s robust performance in encoding galaxy spectra and predicting redshifts across the BGS and ELG datasets. We also compare SpecPT's performance with existing methodologies, positioning it as a powerful and versatile foundational model for spectroscopic analysis.

As shown in Figures \ref{fig:bgs_auto} and \ref{fig:elg_auto}, SpecPT excels at reconstructing spectra, capturing critical features such as emission lines, absorption lines, and the continuum shape. Additionally, it reduces noise and avoids reconstructing artifacts, such as skylines, which might otherwise be misidentified as spectral features. This capability underscores SpecPT’s ability to distinguish intrinsic spectral information from noise, similar to the way an expert astronomer analyzes spectra. However, unlike manual analyses, which are time-intensive and limited in scale, SpecPT can process hundreds of thousands of spectra in minutes, significantly accelerating scientific workflows.

When comparing SpecPT’s redshift measurement performance with prior methodologies, its advancements become evident. Previous works such as \cite{zhou2021spectroscopic} and \cite{stivaktakis2019convolutional} trained CNNs on simulated spectroscopic data for various instruments, reporting variable performance depending on training set size and data signal-to-noise ratio (SNR). While direct comparison is challenging due to differences in datasets, SpecPT’s performance—demonstrated through low NMAD values and tight residual distributions—appears superior to these methods. This comparison reinforces SpecPT’s robustness and adaptability in handling diverse spectroscopic challenges.

SpecPT’s performance is further validated by its alignment with other state-of-the-art transformer-based models designed for spectroscopic redshift measurements. For example, in \cite{parker2024astroclip}, the authors evaluate the AstroCLIP, Spectrum Encoder, and SPENDER \citep{melchior2023autoencoding} models on DESI data in the $0 < z < 0.6$ range, reporting $R^2$ scores of 0.98 to 0.99. The $R^2$ score, also known as the coefficient of determination, is a statistical measure that quantifies how well the predicted values of a model approximate the true values. Mathematically, it is defined as:

\begin{equation}
R^2=1-\frac{\sum\left(z_{\text {true }}-z_{\text {predicted }}\right)^2}{\sum\left(z_{\text {true }}-\bar{z}_{\text {true }}\right)^2}\mcomma
\end{equation}

\noindent where $z_{true}$ are the true redshifts, $z_{predicted}$ are the predicted redshifts, and $\bar{z}_{true}$ is the mean of the true redshifts. An $R^2$ score of 1 indicates perfect predictions, while a score closer to 0 implies that the model performs no better than simply using the mean of the true values. SpecPT achieves an $R^2$ of 0.99 on the BGS dataset, matching the best-performing models in \cite{parker2024astroclip}. Moreover, SpecPT’s redshift residual distributions show similar trends, with errors decreasing in regions of high data density and increasing where data are sparse. These findings highlight SpecPT’s capacity to compete with the most advanced transformer-based models currently available.

One of SpecPT’s most significant accomplishments is its ability to generalize to high-redshift datasets, as demonstrated by its performance on the ELG data. While other studies (such as the ones discussed in this paper) focus on low-redshift ranges, SpecPT extends these capabilities to higher redshifts without a significant drop in performance. The NMAD and outlier fraction values for the ELG dataset, 0.0008 and 0.80\%, respectively, remain comparable to those for the BGS dataset, demonstrating SpecPT’s versatility and reliability across redshift ranges. This consistency highlights its potential for analyzing diverse spectroscopic datasets without specialized retraining for different redshift regimes.

Further supporting this conclusion are the UMAP visualizations of SpecPT’s latent space embeddings, as shown in Figures \ref{fig:bgs_umap} and \ref{fig:elg_umap}. These plots illustrate a clear gradient in redshift, with similar spectra grouped together naturally. The ability of the model to organize spectra in this way reinforces its capability to extract meaningful, interpretable representations of spectroscopic data. This ability of SpecPT would be particularly valuable for tasks beyond redshift measurement, such as identifying outliers or characterizing interstellar medium (ISM) properties such as SFR, Z and gas pressure to name a few.

A notable observation from the results is the complementary behavior of SpecPT across the two datasets. For the BGS dataset, SpecPT performs best at low redshifts, reflecting the dense distribution of data in this range. Conversely, for the ELG dataset, performance improves at higher redshifts, where data density is higher. This suggests that a combined training set incorporating both datasets could produce a model capable of adapting seamlessly across the full redshift range, enhancing its utility for comprehensive spectroscopic surveys.

\section{Summary and Conclusion}
\label{sec:summary}

In this paper, we introduced SpecPT, a novel transformer-based architecture for spectroscopic data analysis, with a focus on autoencoding spectra and measuring redshifts. We demonstrated SpecPT’s capabilities on the BGS and ELG datasets from the DESI EDR. SpecPT successfully performed two key tasks: encoding galaxy spectra through an autoencoder and predicting redshifts with high precision. By leveraging the inherent features of galaxy spectra, SpecPT achieved competitive results compared to state-of-the-art methods, setting the stage for its potential as a foundational model for spectroscopic analysis.

\noindent\textbf{Key Results and Insights:}

\begin{itemize}
    \item \textbf{Spectral Reconstruction}: SpecPT’s autoencoder effectively captured key spectral features, reduced noise, and avoided artifacts like skylines, demonstrating its ability to generalize across diverse datasets.

    \item \textbf{Redshift Measurement}: SpecPT achieved-
    \begin{itemize}[label=\textasteriskcentered]
        \item NMAD values of 0.0006 and 0.0008 for the BGS and ELG datasets, respectively.
        \item Outlier fractions of 0.20\% for BGS and 0.80\% for ELG, reflecting robust performance across different redshift ranges.
    \end{itemize}

    \item \textbf{Complementary Performance}: SpecPT showed complementary strengths across BGS (low redshifts) and ELG (higher redshifts), highlighting the potential benefit of combining datasets to improve performance across the full redshift range.

    \item \textbf{Future Applications}: SpecPT’s latent space learning positions it for tasks beyond redshift measurement, such as ISM property estimation, outlier detection, and galaxy classification.
\end{itemize}

A key observation from this work is the complementary performance of SpecPT across the BGS and ELG datasets. The evolution of redshift residuals with redshift suggests that combining these datasets to create a more balanced training set could enhance SpecPT’s performance across the full redshift range. This is a promising direction for future work, as we plan to explore the creation of a comprehensive training dataset combining BGS, ELG, LRG, and QSO spectra. Such a dataset would provide a balanced redshift distribution, enabling SpecPT to adapt seamlessly to both low- and high-redshift data. This will form the basis of the next paper in this series.

Future work will focus not only on enhancing SpecPT's performance on DESI data but also on extending its application to other spectroscopic datasets, including observations from both ground-based and space-based instruments at higher redshifts. By fine-tuning SpecPT on these datasets, we aim to expand its applicability to data from both space- and ground-based instruments, enabling cross-survey compatibility and broader scientific utility.

Beyond redshift measurement, SpecPT’s ability to learn meaningful latent representations positions it as a multipurpose tool for spectroscopy. Its robust latent space could facilitate tasks such as outlier detection, ISM property estimation, and classification of galaxy types. These capabilities further reinforce its potential to serve as a foundational model for spectroscopy, capable of addressing a wide range of astrophysical challenges.

It is also important to emphasize that the DESI Early Data Release (EDR) used in this work represents only 2\% of the full DESI dataset. Training SpecPT on the complete DESI dataset in the future is expected to significantly enhance its performance, enabling it to tackle even larger datasets, such as the COSMOS Spectroscopic Archive. Expanding SpecPT’s training base will not only improve its accuracy but also further cement its position as a foundational model capable of scaling to the demands of next-generation spectroscopic surveys.

In conclusion, SpecPT lays the groundwork for a multipurpose spectroscopic analysis framework. While this paper focuses on redshift measurement, SpecPT’s flexibility and scalability make it well-suited for a variety of spectroscopic tasks. The continued evolution of SpecPT in future work will aim to unlock its full potential, transforming how spectroscopic data is analyzed and interpreted. This paper is the first in a series of studies that will refine and expand SpecPT’s capabilities, ultimately paving the way for transformative advancements in spectroscopic science.

\section{Acknowledgments}

The authors acknowledge Research Computing at the Rochester Institute of Technology \citep{researchcomprit} for providing the computational resources and support that were instrumental in achieving the research results reported in this paper. We also extend our gratitude to Dr. Swartzlander for his support, with partial computational resources provided under the U.S. ONR grant N00014-23-1-2513.

This material is based upon work supported by the National Science Foundation under Grant No. 2009572 and NASA under award No. 80NSSC22K0854.

RP would also like to thank Roberto Ceralo, Tetsushi Yamada, and Pranshu Malviya for their insightful and engaging discussions.

\bibliography{references}{}
\bibliographystyle{aasjournal}
\end{document}